\documentclass[sigconf,table,natbib=false]{acmart}
\usepackage[
	backend=biber,
	style=numeric,
	firstinits=true,
	url=false,
	isbn=false,
	]{biblatex}
\usepackage{graphicx}

\setcopyright{none}

\begin{document}

\title[Comparing Encodings of Natural and Mathematical Language]{Classification and Clustering\\ of arXiv Documents, Sections, and Abstracts,\\ Comparing Encodings of\\ Natural and Mathematical Language}

\author{Philipp Scharpf}
\affiliation{
  \institution{University of Konstanz, Germany}
}
\email{philipp.scharpf@uni-konstanz.de}

\author{Moritz Schubotz}
\affiliation{
  \institution{University of Wuppertal and FIZ Karlsruhe, Germany}
}
\email{moritz.schubotz@fiz-karlsruhe.de}

\author{Abdou Youssef}
\affiliation{
  \institution{George Washington University, United States}
}
\email{ayoussef@gwu.edu}

\author{Felix Hamborg}
\affiliation{
  \institution{University of Konstanz, Germany}
}
\email{felix.hamborg@uni-konstanz.de}

\author{Norman Meuschke}
\affiliation{
  \institution{University of Wuppertal \& University of Konstanz, Germany}
}
\email{meuschke@uni-wuppertal.de}

\author{Bela Gipp}
\affiliation{
  \institution{University of Wuppertal \& University of Konstanz, Germany}
}
\email{gipp@uni-wuppertal.de}

\renewcommand{\shortauthors}{Scharpf et al.}


\copyrightyear{2020}
\acmYear{2020}
\setcopyright{acmcopyright}
\acmConference[JCDL '20] {ACM/IEEE Joint Conference on Digital Libraries in 2020}{August 1--5, 2020}{Virtual Event, China}
\acmPrice{15.00}
\acmDOI{10.1145/3383583.3398529}
\acmISBN{978-1-4503-7585-6/20/08}


\begin{abstract}
In this paper, we show how selecting and combining encodings of natural and mathematical language affect classification and clustering of documents with mathematical content. We demonstrate this by using sets of documents, sections, and abstracts from the arXiv preprint server that are labeled by their subject class (mathematics, computer science, physics, etc.) to compare different encodings of text and formulae and evaluate the performance and runtimes of selected classification and clustering algorithms. Our encodings achieve classification accuracies up to $82.8\%$ and cluster purities up to $69.4\%$ (number of clusters equals number of classes), and $99.9\%$ (unspecified number of clusters) respectively. We observe a relatively low correlation between text and math similarity, which indicates the independence of text and formulae and motivates treating them as separate features of a document. The classification and clustering can be employed, e.g., for document search and recommendation. Furthermore, we show that the computer outperforms a human expert when classifying documents. Finally, we evaluate and discuss multi-label classification and formula semantification.
\end{abstract}

\maketitle
\section{Introduction}\label{sec:Intro}

The computational analysis of documents (e.g., for Recommender systems) from Science, Technology, Engineering and Mathematics (STEM) is particularly challenging since it involves both Natural Language Processing (NLP) and Mathematical Language Processing (MLP) to simultaneously investigate text and formulae. While NLP already relies heavily on Machine Learning (ML) techniques, their use in MLP is still being explored. In this paper, we show how methods of NLP and MLP can be combined to enable the use of ML in Information Retrieval (IR) applications on documents with mathematical content.
Machine Learning (ML) has been evolving since Alan Turing's proposal of a \textit{Learning Machine}~\cite{turing1950computing}. It has decisively promoted fields such as Computer Vision, Speech Recognition, Natural Language Processing, and Information Retrieval, with a vast number of applications, e.g., in Medical Diagnosis, Financial Market Analysis, Fraud Detection, Recommender Systems, Object Recognition, and Machine Translation.

Natural Language Processing (NLP) is an interdisciplinary field involving both computer science and linguistics to develop methods that enable computers to process and analyze natural language data~\cite{nadkarni2011natural}. Originally evolving from automatic translation - Georgetown experiment~\cite{DBLP:conf/amta/Hutchins04}, and chatbots - ELIZA~\cite{DBLP:journals/cacm/Weizenbaum66} - the discipline has made fast advancements in Part-of-speech (POS) tagging, Named Entity Recognition (NER) and Relationship extraction~\cite{DBLP:conf/pakdd/WongLB09}.
Recently, NLP has especially been enriched by enhanced Deep Learning capabilities~\cite{DBLP:journals/cim/YoungHPC18}.
Mathematical Language Processing (MLP) was first coined and introduced by Pagel and Schubotz~\cite{DBLP:conf/mkm/PagelS14} as a term and discipline that is concerned with analyzing mathematical formulae, analogous to how NLP is dealing with natural language sentences. This comprises the semantic enrichment of mathematical formulae and their constituents to automatically infer their meaning from the context~\cite{DBLP:conf/sigir/SchubotzGLCMGYM16} (surrounding text, mathematical topic or discipline, etc.).

All these research fields and disciplines are joint contributors in the analysis of documents with both natural and mathematical language (containing text and formulae). This paper illustrates the synergy of NLP and MLP in ML applications. We employed a set of 4900 documents, 3500 sections, and 1400 abstracts from the arXiv preprint server (\url{arxiv.org}) that are labeled by their subject class (mathematics, computer science, physics, etc.) to compare different encodings (doc2vec, tf-idf) of text and formulae. We evaluated the performance and runtimes of selected classification and clustering algorithms, observing classification accuracies up to $82.8\%$ and cluster purities up to $69.4\%$ for a fixed number of clusters and $99.9\%$ if the number of clusters is unspecified.

The paper is structured as follows. We first review related work in NLP and MLP, followed by the current state of research in text and math classification and clustering in Section \ref{sec:RelWork}.
In the main Section \ref{sec:OurStudy}, we describe our experimental setting - the employed datasets, data cleaning, encoding types, classification and clustering algorithms, and evaluation measures. In the subsequent Section \ref{sec:Results}, we present the obtained accuracies/purities of different classification/clustering algorithms operating on various text or math encodings of the mathematical documents. Finally, we conclude our study and outline some future questions and potentials in Section \ref{sec:ConclOutl}.

\section{Related Work}\label{sec:RelWork}

\subsection{Natural Language Processing}

Since Machine Learning methods require a formal (abstract mathematical) representation, natural language has to be converted into word vectors. This is typically done via Word2Vec~\cite{DBLP:journals/corr/abs-1301-3781} using Bag-of-Words (BOW) or Skip-Grams (SG), or encoding term-frequency (tf) or term frequency-inverse document frequency (tf-idf).
Whole documents or document sections can be represented, i.a., by Doc2Vec~\cite{DBLP:conf/icml/LeM14} features, which are learned using (Deep) Neural Networks~\cite{DBLP:journals/cim/YoungHPC18}. Vectors of words or documents generated by Word2Vec or Doc2Vec were observed to be semantically close with respect to algebraic distance metrics and can be used to reconstruct linguistic contexts of words~\cite{DBLP:conf/nips/MikolovSCCD13}. This enables comparisons of the semantic content of documents, e.g., for Recommender systems or text-topic classification.

\subsection{Mathematical Language Processing}

The Mathematical Language Processing (MLP) project~\cite{DBLP:conf/mkm/PagelS14} was introduced as an attempt to disambiguate identifiers occurring in mathematical formulae. Retrieving the natural language definition of the identifiers using Part-Of-Speech (POS) tags combined with numerical statistics for the candidate ranking yielded high accuracies of around 90\%.

The Part-Of-Math (POM) tagging project~\cite{DBLP:conf/mkm/Youssef17} also aims at math disambiguation and math semantics determination for the enrichment of math expressions. Scanning a math input document, definitive tags (operation, relation, etc.) and tentative features (alternative roles and meanings) are assigned to math expressions using a semantic database that was created for the project.

Furthermore, a comprehensive Math Knowledge Processing project was started~\cite{DBLP:conf/mkm/YoussefM18} to explore sequence-to-sequence translation from LaTeX typesetting to MathML markup using Math2Vec encodings of the formulae. A dataset of 6000 papers has been collected to be used as training and testing data for the semantification of the identifiers.

\subsection{Classification}

\paragraph{Text Document Classification}

Automatic document classification (ADC) has increasingly gained interest due to the vast availability of documents needing to be rapidly categorized. Advantages over the knowledge engineering approach of "manual" labeling by domain experts are efficiency (saving time) and easy portability (general techniques). Compared to traditional methods, e.g., fuzzy logic, ML methods are less interpretable, but often more effective and thus widely used. One differentiates single-label vs. multi-label, as well as hard (top one) vs. ranking text categorization~\cite{DBLP:journals/csur/Sebastiani02}. Applications of ADC include spam filtering, sentiment analysis, product categorization, speech categorization, author and text genre identification, essay grading, automatic document indexing, word sense disambiguation, and hierarchical categorization of web pages~\cite{DBLP:journals/eswa/MironczukP18}.

\paragraph{Mathematical Document Classification}

For the classification of mathematical documents, a mathematics-aware Part Of Speech (POS) tagger was developed to extend the dictionaries for keyphrase identification via noun phrases (NPs) by symbols and mathematical formulae~\cite{DBLP:conf/mkm/SchonebergS14}. The aim was to aid mathematicians in their search for relevant publications by classified tags, such as `named mathematical entities' where, e.g., names of mathematicians indicate being potential parts of names for a special conjecture, theorem, approach or method. The hierarchical Mathematics Subject Classification~\cite{DBLP:conf/icms/Dong18} scheme was employed.

Mathematical formulae (available as TeX code) were transformed to unique but random character sequences, e.g. $x?(t)=f(t,x(t))$ to "kqnompjyomsqomppsk". Prior to the classification, key NP candidates were extracted from the full text or abstract and evaluated by experts (editors or reviewers removing, changing or adding phrases). The authors suggest that for scalable automatic extraction of key phrases, titles and abstracs are more accessible and suitable because they already summarize the publication content. The developed tools were tested for key phrase extraction and classification in the database zbMATH~\cite{zbMATH}, obtaining best results using an SVM sequential minimal optimization algorithm with polynomial kernel. For 26 of the 63 top-level classes the precision was higher than 0.75 and only for 4 classes smaller than 0.5. Controversial criteria such as quality, correctness, completeness, uncertainty, subjectivity, reliability were discussed.

\subsection{Clustering}

\paragraph{Text Document Clustering}

Motivated by the need for unstructured document organization, summarization, and knowledge discovery, clustering methods are increasingly used for efficient representation and visualization. Applications of clustering in science and business include search engines, recommender systems, duplicate and plagiarism detection, and topic modeling. Similar documents are grouped such that intra-class similarities are high, while inter-class similarity is low~\cite{shah2012document}.
Being an unsupervised ML method, clustering does not need any prior knowledge about the class distribution, at the cost of the results potentially not being properly understandable or interpretable for humans. Distinctions are made between hard (disjoint) vs. soft (overlapping), as well as agglomerative (bottom-up) vs. divisive (top-down) clustering. In a survey on semantic document clustering~\cite{naik2015survey}, augmentation by synonyms and domain specific ontologies is suggested to improve Latent Semantic Analysis (LSA), and Word Sense Disambiguation (WSD). Among the challenges of text clustering are the extraction and selection of appropriate features, similarity measure, clustering method and algorithm, efficient implementation, meaningful cluster labeling, and appropriate evaluation criteria. A review of the history and methods of document clustering can be found at~\cite{premalatha2010literature}.

\paragraph{Mathematical Formula Clustering}

First investigations of how clustering algorithms perform on mathematical formulae have been made by two groups~\cite{ma2010feature, adeel2012efficient}. The first group compared three clustering algorithms - K-Means, Agglomerative Hierarchical Clustering (AHC), and Self Organizing Map (SOM) - on 20 training and 20 test samples, showing Top-5 and Top-10 accuracies between 82\% and 99\% with the discovery that all three achieved similarly high results. The second group aimed at the speedup
of formula search, which is why they also discussed the runtimes of the algorithms with the observation that K-means outperformed the other two (Self Organizing Map, Average-link) with 96\% precision. For further details, the reader is referred to the respective publications.
Clustering-based retrieval of mathematical formulae can have several applications. It can speed up formula search, e.g., on the Digital Library of Mathematical Functions (DLMF)~\cite{DBLP:journals/amai/Lozier03} or the arXiV.org e-Print archive~\cite{mckiernan2000arxiv} by grouping indexed formulae or documents. Given a formula search query, the closest cluster centroid is determined first, reducing the remaining search space by a factor that is the cluster parameter $k$. Furthermore, clustering possible solutions to mathematical exercises can help in automatic grading and feedback for learners assignments~\cite{DBLP:conf/lats/LanVWB15}.

\section{Our Study}\label{sec:OurStudy}

In this section, we investigate how Machine Learning (ML) can combine Natural Language Processing (NLP) and Mathematical Language Processing (MLP) when classifying and clustering documents (docs), sections (secs), and abstracts (abs) containing both text and formulae.

Our \textbf{research} was driven by the following \textbf{questions}:

1) \textit{How does selecting and combining encodings of natural and mathematical language affects classification (accuracy) and clustering (purity) of documents with mathematical content?}

2) \textit{Which encoding (content=text/math, method=2vec/tf-idf) or algorithm (classification/clustering) has the highest performance (accuracy/purity) and shortest runtime?}

The following section starts by presenting the employed datasets, followed by a description of our data extraction pipeline, and the encodings. We then report an examination of the correlation between text and math similarity, followed by our investigation to classify and cluster STEM docs, secs, and abs from the arXiv preprint server by their subject class (mathematics, computer science, physics, etc.) using the contained text and formulae. Finally, we compare the classification confusion of the computer to a human expert, summarize our findings and outline some future directions and experiments.

We conducted experiments using 9800 samples 
(documents, sections, abstracts) with 400 different settings (encodings, methods, algorithms).

\textbf{Our code is available at \url{https://purl.org/class_clust_arxiv_code}.}

\subsection{Datasets}

\paragraph{SigMathLing arXMLiv-08-2018}

Provided by the Special Interest Group on Maths Linguistics (\url{sigmathling.kwarc.info}), the arXMLiv-08-2018 dataset contains 
137864 HTML document files 
(\url{w3c.org/html}).
We selected an equal distribution of the first 350 documents from each of the following subject classes: ['hep-ph', 'astro-ph', 'quant-ph', 'physics', 'cond-mat', 'hep-ex', 'hep-lat', 'nucl-th', 'nucl-ex', 'hep-th', 'math', 'gr-qc', 'nlin', 'cs'], yielding a total of $14 \times 350 = 4900$ documents.

\paragraph{NTCIR-11/12 MathIR arXiv}

Provided by the National Institute of Informatics Testbeds and Community for Information Access Research Project (NTCIR)~\cite{DBLP:conf/ntcir/AizawaKOS14, DBLP:conf/ntcir/ZanibbiAKOTD16}, the MathIR arXiv dataset contains 
104062 TEI document section files (\url{www.tei-c.org}).
We selected an equal distribution of the first 250 sections and 100 abstracts from each of the following subject classes: ['astro-ph', 'cond-mat', 'cs', 'gr-qc', 'hep-lat', 'hep-ph', 'hep-th', 'math-ph', 'math', 'nlin', 'quant-ph', 'physics', 'alg-geom', 'q-alg'], yielding totals of $14 \times 250 = 3500$ sections and $14 \times 100 = 1400$ abstracts.

\subsection{Data Extraction}

\paragraph{Text}

From the HTML documents and TEI section files, we retrieved the textual content using the \textit{nltk}~\cite{DBLP:conf/acl/ManningSBFBM14} \textit{RegexpTokenizer} and \textit{corpus English stopword set}. We cleaned the raw text strings by lowering and removing stopwords, mathematical formulae, digits and words with less than three characters. The cleaning increased classification accuracies up to a factor of 3.35 for tf-idf encodings while achieving less improvements for doc2vec encodings.

\paragraph{Formulae}

We retrieved the mathematical content using the Python package BeautifulSoup~\cite{DBLP:journals/jcp/ZhengHP15}. We isolated the formulae from \verb|<formula>| (TEI) and \verb|<math>| (HTML) tags, and extracted the operators and identifiers and from \verb|<mo>| and \verb|<mi>| tags respectively. The formulae were converted from TeX/LaTeX format to XML and and HTML5 with MathML (\url{w3c.org/Math}) via LaTeXML~\cite{DBLP:conf/mkm/GinevSMK11}.

\subsection{Encodings}

We encoded the retrieved text and formulae using the \textit{TfidfVectorizer} from the Python package \textit{Scikit-learn}~\cite{DBLP:journals/jmlr/PedregosaVGMTGBPWDVPCBPD11} and \textit{Doc2Vec} model~\cite{DBLP:conf/icml/LeM14} from the Python package \textit{Gensim}~\cite{vrehuuvrek2011scalability}.

After creating a LabeledLineSentence iterator for the vocabulary, the model was built with size=300, window=10, min\_count=5, workers=11, alpha=0.025, min\_alpha=0.025, iter=20 and trained 10 epochs with model.alpha-=0.002.

Table \ref{tab:Enc} lists the encodings that deserve further explanation.
The \textit{surroundings} encoding uses text surroundings (within +-500 characters, excluding stopwords and letters) of single identifiers as their putative meanings.

\begin{table}[h]
\centering
\caption{Special encodings of mathematical formula content with explanation.}
\label{tab:Enc}
\vspace{5pt}
\begin{tabular}{|l|l|}
\hline
\textbf{Encoding}        & \textbf{Explanation} \\ \hline
Math\_op            & formula operators (+,-, etc.)     \\ \hline
Math\_id            & formula identifiers (x,y,z, etc.)     \\ \hline
Math\_opid          & formula operators and identifiers    \\ \hline
Math\_surroundings     & text surroundings of identifiers     \\ \hline
\end{tabular}
\end{table}

Summing up, we varied the following experimental parameters: 1) encoded data types or batch size (documents, sections, abstracts, summarizations), 2) encoded data features (text, math), 3) type of math encoding (op, id, opid, surroundings), 4) encoding method (doc2vec, tf-idf), 5) classification/clustering algorithm (performance, runtime).

\subsection{Correlation between Text and Math Similarity}

First, we determined the correlation of the cosine similarity (inner product) between text and math encodings. For each document, section or abstract, we calculated the similarities with all the other docs/secs/abs in both text and math encodings (different vector spaces) seperately. This means, we investigated whether if two documents are similar in their text encoding, they are also similar in their math encoding.
Table \ref{tab:Corr} lists the results of our comparison.
The low correlations indicate that in principle, the independence of text and math encodings leave a potential for improvement of ML algorithms by combining the two, which was explored. Besides, it suggests that in a Recommender System for STEM documents, it will be beneficial to provide the user with weighting parameters for text and math (if relatively uncorrelated), to customize the recommendations.

\begin{table}[h]
\centering
\caption{Correlations between text and math (cosine) similarity of individual documents and sections.}
\label{tab:Corr}
\vspace{5pt}
\begin{tabular}{|l|l|l|}
\hline
Comparison/Domain x              & doc       & sec       \\ \hline
x2vecText - x2vecMath\_op        & 0.14 & 0.16 \\ \hline
x2vecText - x2vecMath\_id        & 0.12 & 0.11 \\ \hline
x2vecText - x2vecMath\_opid      & 0.16 & 0.15 \\ \hline
x2vecText - x2vecMath\_surroundings & 0.21 & 0.27 \\ \hline
\end{tabular}
\end{table}

\subsection{Classification and Clustering}

Using the selection of 4900 documents, 3500 sections, and 1400 abstracts from the arXiv, we compared the influence of text and formulae on the performance of a subject class ['math', 'physics', 'cs', etc.] classification. We subsequently clustered the doc/sec/abs vectors;
for the {\it KMeans, Agglomerative}, and {\it GaussianMixture} cluterers, we fixed the number of clusters to 14 (= the number of
labeled classes), while for the {\it Affinity, MeanShift}, and {\it HDBSCAN} clusterers, no number of clusters was fixed. The encodings secText\_tfidf and sec2vecMath\_surroundings needed a PCA dimensionality reduction before the clustering with MeanShift and GaussianMixture was possible.

\subsection{Evaluation}

We used 10-fold cross-validation\footnote{We observed that the split k had only a small impact on the result.}, while comparing the accuracy, purity, and relative runtimes of selected single or ensemble classifiers and clustering algorithms (with their respective default metrics) with or without fixed cluster number, provided by the Python package \textit{Scikit-learn}~\cite{DBLP:journals/jmlr/PedregosaVGMTGBPWDVPCBPD11}.

For the classification, we calculated the accuracy as the number of correctly classified samples divided by the sample size and averaged over all splittings of the cross-validation.

For the clustering, we compared the clusters to the labeled classes, calculating the cluster purity as the number of data points of the class that makes up the largest fraction of the cluster divided by the cluster size and averaged over all clusters.

\section{Results} \label{sec:Results}

The results of the classification and clustering are shown in Tables \ref{tab:Class} and \ref{tab:Clust}.

In contrast to the classification, the clustering of math vectors yielded partly better results than the text clustering. A combination of both text and math yielded no significant improvement over the separate encodings.

\subsection{Classification}

\begin{table*}[ht]
\caption{Classification accuracies of 4900 arXiv \textbf{doc}uments (above), 3500 \textbf{sec}tions (middle), and 1400 \textbf{abs}tracts (below) into 14 subject classes using different classifier (columns), and text or math encodings (rows). The highest mean/maximum is highlighted in yellow/red. It is orange if an encoding or classifier yields both the highest mean and maximum value. The shortest relative runtime is marked in green.}
\vspace{10pt}
\resizebox{\textwidth}{!}{%
\begin{tabular}{|l|l|l|l|l|l|l|l|l|l|l|}
\hline
\textbf{Encoding/Classifier}        & \textbf{LogReg} & \cellcolor{red}\textbf{LinSVC} & \textbf{RbfSVC} & \textbf{kNN}   & \cellcolor{yellow}\textbf{MLP}      & \textbf{DecTree} & \textbf{RandForest} & \textbf{GradBoost} & \textbf{Mean}    & \textbf{Max}      \\ \hline
\textbf{doc2vecText}                & 64.5   & 60.3   & 75.2   & 18.6  & 72.8     & 31.7    & 45.6       & 64.6      & 54.2    & 75.2     \\ \hline
\cellcolor{orange}\textbf{docText\_tfidf}             & 80.6   & 82.8   & 72.8   & 78.1  & 82.6     & 57.0    & 51.8       & 75.8      & \cellcolor{yellow}72.7    & \cellcolor{red}82.8     \\ \hline
\textbf{doc2vecMath\_op}            & 37.7   & 37.4   & 38.1   & 22.8  & 31.8     & 16.1    & 21.3       & 33.8      & 29.9    & 38.1     \\ \hline
\textbf{docMath\_op\_tfidf}         & 14.9   & 15.0   & 13.7   & 10.1  & 14.7     & 14.4    & 14.6       & 14.7      & 14.0    & 15.0     \\ \hline
\textbf{doc2vecMath\_id}            & 42.2   & 36.0   & 33.8   & 22.5  & 43.3     & 14.8    & 17.8       & 36.8      & 30.9    & 43.3     \\ \hline
\textbf{docMath\_id\_tfidf}         & 23.0   & 22.8   & 16.4   & 15.2  & 23.0     & 18.0    & 20.9       & 22.1      & 20.2    & 23.0     \\ \hline
\textbf{doc2vecMath\_opid}          & 45.7   & 39.7   & 24.8   & 17.7  & 46.7     & 14.2    & 17.9       & 39.6      & 30.8    & 46.7     \\ \hline
\textbf{docMath\_opid\_tfidf}       & 25.5   & 25.8   & 17.1   & 16.7  & 25.2     & 17.3    & 21.4       & 24.7      & 21.7    & 25.8     \\ \hline
\textbf{doc2vecMath\_surroundings}     & 49.5   & 46.5   & 24.8   & 14.4  & 51.3     & 12.0    & 14.8       & 40.6      & 31.7    & 51.3     \\ \hline
\textbf{docMath\_surroundings\_tfidf}  & 63.2   & 63.4   & 43.7   & 7.2   & 64.7     & 37.1    & 44.0       & 56.4      & 47.5    & 64.7     \\ \hline
\textbf{doc2vecTextMath\_opid}      & 64.2   & 59.3   & 74.7   & 10.4  & 71.7     & 31.8    & 41.5       & 63.3      & 52.1    & 74.7     \\ \hline
\textbf{doc2vecTextMath\_surroundings} & 61.7   & 57.4   & 61.4   & 23.7  & 70.6     & 31.6    & 41.4       & 64.0      & 51.5    & 70.6     \\ \hline
\textbf{Mean}                       & 47.7   & 45.5   & 41.4   & 21.5  & \cellcolor{yellow}49.9     & 24.7    & 29.4       & 44.7      & 38.1    & 49.9     \\ \hline
\textbf{Max}                        & 80.6   & \cellcolor{red}82.8   & 75.2   & 78.1  & 82.6     & 57.0 
& 51.8       & 75.8      & 73.0    & 82.8     \\ \hline
\textbf{Runtime {[}\%{]}}          & 1.1    & 1.6    & 4.6    & 0.1   & 100.0    & 0.4     & \cellcolor{green}0.1        & 46.3      & 19.3    & 100.0    \\ \hline
\end{tabular}}
\\[30pt]
\resizebox{\textwidth}{!}{%
\begin{tabular}{|l|l|l|l|l|l|l|l|l|l|l|}
\hline
\textbf{Encoding/Classifier}        & \textbf{LogReg}  & \textbf{LinSVC}  & \textbf{RbfSVC}  & \textbf{kNN}  & \cellcolor{orange}\textbf{MLP}     & \textbf{DecTree} & \textbf{RandForest} & \textbf{GradBoost} & \textbf{Mean}    & \textbf{Max}     \\ \hline
\textbf{sec2vecText}                & 51.7    & 51.9    & 51.8    & 58.0 & 61.9    & 51.8    & 33.5       & 51.7      & 51.5    & 61.9    \\ \hline
\cellcolor{orange}\textbf{secText\_tfidf}             & 68.9    & 69.0    & 69.4    & 70.1 & 77.5     & 69.2    & 49.3       & 69.1      & \cellcolor{yellow}67.8    & \cellcolor{red}77.5    \\ \hline
\textbf{sec2vecMath\_op}            & 29.9    & 30.2    & 29.9    & 21.2 & 40.2    & 30.2    & 18.5       & 29.9      & 28.8    & 40.2    \\ \hline
\textbf{secMath\_op\_tfidf}         & 14.8    & 14.6    & 14.5    & 11.2 & 16.7    & 14.7    & 13.7       & 14.8      & 14.4    & 16.7    \\ \hline
\textbf{sec2vecMath\_id}            & 27.6    & 28.0    & 27.9    & 14.0 & 40.7    & 28.2    & 13.8       & 28.0      & 26.0    & 40.7    \\ \hline
\textbf{secMath\_id\_tfidf}         & 23.8    & 23.9    & 23.7    & 16.1 & 24.1    & 23.9    & 20.3       & 23.9      & 22.5    & 24.1    \\ \hline
\textbf{sec2vecMath\_opid}          & 30.1    & 30.2    & 30.0    & 10.6 & 42.3    & 29.9    & 16.5       & 30.3      & 27.5    & 42.3    \\ \hline
\textbf{secMath\_opid\_tfidf}       & 27.0    & 27.1    & 26.1    & 17.3 & 25.9    & 26.6    & 22.6       & 26.5      & 24.9    & 27.1    \\ \hline
\textbf{sec2vecMath\_surroundings}     & 32.7    & 33.3    & 32.3    & 10.3 & 47.5    & 32.5    & 12.3       & 32.3      & 29.2    & 47.5    \\ \hline
\textbf{secMath\_surroundings\_tfidf}  & 54.6    & 54.5    & 55.0    & 8.1  & 61.0    & 55.3    & 41.9       & 54.7      & 48.1    & 61.0    \\ \hline
\textbf{\textbf{sec2vecText}Math\_opid}      & 49.9    & 50.4    & 50.2    & 52.0 & 63.0    & 50.3    & 31.0       & 50.5      & 49.7    & 63.0    \\ \hline
\textbf{\textbf{sec2vecText}Math\_surroundings} & 50.8    & 50.7    & 50.7    & 23.7 & 60.5    & 50.8    & 31.8       & 50.8      & 46.2    & 60.5    \\ \hline
\textbf{Mean}                       & 38.5    & 38.7    & 38.5    & 26.1 & \cellcolor{yellow}46.8    & 38.6    & 25.4       & 38.5      & 36.4    & 46.8    \\ \hline
\textbf{Max}                        & 68.9    & 69.0    & 69.4    & 70.1 & \cellcolor{red}77.5    & 69.2    & 49.3       & 69.1      & 67.8    & 77.5    \\ \hline
\textbf{Runtime {[}\%{]}}          & 100.0   & 100.0   & 95.6    & 0.2  & 78.5    & 100.0   & \cellcolor{green}0.2        & 100.0     & 71.8    & 100.0   \\ \hline
\end{tabular}}
\\[30pt]
\resizebox{\textwidth}{!}{%
\begin{tabular}{|l|l|l|l|l|l|l|l|l|l|l|}
\hline
\textbf{Encoding/Classifier}  & \textbf{LogReg} & \textbf{LinSVC} & \textbf{RbfSVC} & \textbf{kNN}  & \cellcolor{orange}\textbf{MLP}    & \textbf{DecTree} & \textbf{RandForest} & \textbf{GradBoost} & \textbf{Mean}  & \textbf{Max}    \\ \hline
\textbf{abs2vecText}          & 42.6   & 38.5   & 50.2   & 25.6 & 47.1   & 17.1    & 21.4       & 38.1      & 35.1  & 50.2   \\ \hline
\cellcolor{orange}\textbf{absText\_tfidf}       & 58.9   & 61.1   & 49.1   & 50.9 & 61.6   & 33.1    & 37.4       & 46.4      & \cellcolor{yellow}49.8  & \cellcolor{red}61.6   \\ \hline
\textbf{abs2vecMath\_opid}    & 26.1   & 26.5   & 22.1   & 16.6 & 23.8   & 13.8    & 17.0       & 22.0      & 21.0  & 26.5   \\ \hline
\textbf{absMath\_opid\_tfidf} & 10.6   & 10.7   & 9.9    & 8.4  & 10.4   & 10.5    & 10.2       & 10.5      & 10.2  & 10.7   \\ \hline
\textbf{Mean}                 & 34.6   & 34.2   & 32.8   & 25.4 & \cellcolor{yellow}35.7   & 18.6    & 21.5       & 29.3      & 29.0  & 35.7   \\ \hline
\textbf{Max}                  & 58.9   & 61.1   & 50.2   & 50.9 & \cellcolor{red}61.6   & 33.1    & 37.4       & 46.4      & 50.0  & 61.6   \\ \hline
\textbf{Runtime {[}\%{]}}    & 1.8    & 8.7    & 3.5    & \cellcolor{green}0.2  & 100.0  & 1.6     & 0.4        & 55.0      & 21.4  & 100.0  \\ \hline
\end{tabular}}
\\[40pt]
\label{tab:Class}
\end{table*}

Table \ref{tab:Class} shows the classification accuracies of the 
individual classification algorithms using text or math encodings of the documents, sections, and abstracts.

The best encoding for docs, secs and abs is always Text\_tfidf. The most accurate algorithm is MLP 
(Multilayer Perceptron, with hidden layer size 500), except for docs where LinSVC yields a slightly higher maximum value. The fastest algorithms are kNN and Random Forest. For kNN and DecTree there is a high discrepancy between the values of doc2vecText and docText\_tfidf encodings.
While for text the tf-idf encoding is better, for math it is the doc2vec encoding, with the exception of the surroundings encoding which is, even though connected to mathematical identifiers, effectively text.

An overall comparison of x2vec and tf-idf including both text and math shows that the former outperforms the latter with mean(doc2vecX, sec2vecX, abs2vecX) = (40.2, 37.0, 28.0) mostly greater than mean(doc\_tfidf, sec\_tfidf, abs\_tfidf) = (35.2, 35.5, 30.0), and mean(2vec) = 35.1 $>$ mean(tfidf) = 33.6 summarized. The mean of the means is decaying from docs (38.1) to secs (36.4) to abs (29.0). The surroundings encoding (especially with tf-idf) is better than the math encodings of operators (op), identifiers (id) or both (opid). Given that mean(doc\_text, doc\_math, doc\_textmath) = (63.4, 28.3, 51.8), and mean(sec\_text, sec\_math, sec\_textmath) = (59.7,  27.7, 47.9), it is striking that for the classification, the text encodings yield better results than the standalone math encodings.

We tested some other algorithms that are not listed: Gaussian Naive Bayes yields accuracies of mostly less than 10\%; Multinomial Naive Bayes could not be carried out on the text vectors due to the negative values of their continuous distribution.

\subsection{Clustering}

\begin{table*}[ht]
\caption{Clustering purities of 4900 arXiv \textbf{doc}uments (above), 3500 \textbf{sec}tions (middle), and 1400 \textbf{abs}tracts (below) with 14 subject classes using different clusterers (columns), and text or math encodings (rows). The highest mean/maximum is highlighted in yellow/red for the group of clusterers with specified cluster number (KMeans, Agglomerative, GaussianMixture) and unspecified (Affinity, MeanShift, HDBSCAN) respectively. It is orange if an encoding or clusterer yields both the highest mean and maximum. The shortest relative runtime is marked in green.}
\vspace{10pt}
\resizebox{\textwidth}{!}{%
\begin{tabular}{|l|l|l|l|l|l|l|l|l|}
\hline
\textbf{Encoding/Clusterer}        & \textbf{KMeans} & \cellcolor{yellow}\textbf{Affinity} & \textbf{Agglomerative} & \cellcolor{red}\textbf{MeanShift} & \cellcolor{orange}\textbf{GaussianMixture} & \textbf{HDBSCAN} & \textbf{Mean}   & \textbf{Max}     \\ \hline
\textbf{doc2vecText}                & 57.5   & 73.4     & 57.9          & 7.1       & 56.5            & 77.0    & 54.9   & 85.6    \\ \hline
\cellcolor{yellow}\textbf{docText\_tfidf}             & 63.2   & 83.5     & 64.9          & 83.5       & 55.3             & 87.8    & \cellcolor{yellow}75.3   & 89.5    \\ \hline
\textbf{doc2vecMath\_op}            & 23.3   & 78.2     & 20.9          & 94.6      & 21.0            & 45.0    & 47.2   & 94.6    \\ \hline
\textbf{docMath\_op\_tfidf}         & 34.5   & 95.5     & 35.8          & 82.9      & 45.5            & 46.1    & 56.7   & 82.9    \\ \hline
\textbf{doc2vecMath\_id}            & 41.5   & 81.6     & 29.0          & 99.8      & 68.7            & 43.9    & 60.8   & 99.8    \\ \hline
\textbf{docMath\_id\_tfidf}         & 24.0   & 67.1     & 21.1          & 92.7      & 19.4            & 37.6    & 43.7   & 92.7    \\ \hline
\textbf{doc2vecMath\_opid}          & 51.0   & 46.5     & 36.4          & 97.9      & 42.0            & 49.1    & 53.8   & 97.9    \\ \hline
\textbf{docMath\_opid\_tfidf}       & 18.0   & 76.9     & 18.7          & 7.1       & 25.8            & 34.3    & 30.1   & 57.8    \\ \hline
\cellcolor{red}\textbf{doc2vecMath\_surroundings}     & 62.7   & 96.5     & 33.9          & 99.9      & 69.4            & 7.1     & 61.6   & \cellcolor{red}99.9    \\ \hline
\textbf{docMath\_surroundings\_tfidf}  & 36.0   & 21.8     & 44.6          & 93.6       & 93.4             & 33.1    & 53.8   & 44.6    \\ \hline
\textbf{doc2vecTextMath\_opid}      & 55.9   & 67.5     & 58.5          & 7.1       & 52.5            & 44.1    & 47.6   & 67.5    \\ \hline
\textbf{doc2vecTextMath\_surroundings} & 59.0   & 94.4     & 52.8          & 99.4      & 61.0            & 43.9    & 68.4   & 99.4    \\ \hline
\textbf{Mean}                       & 43.9   & \cellcolor{yellow}73.6     & 39.5          & 72.1      & \cellcolor{yellow}50.9            & 45.8    & 54.3   & 68.9    \\ \hline
\textbf{Max}                        & 63.2   & 96.5     & 64.9          & \cellcolor{red}99.9      & \cellcolor{red}69.4            & 89.5    & 80.6   & 99.9    \\ \hline
\textbf{Runtime {[}\%{]}}          & 5.9    & 3.7      & 7.7           & 100.0     & 2.0             & \cellcolor{green}0.6     & 20.0   & 100.0   \\ \hline
\end{tabular}}
\\[30pt]
\resizebox{\textwidth}{!}{
\begin{tabular}{|l|l|l|l|l|l|l|l|l|}
\hline
\textbf{Encoding/Clusterer}        & \textbf{KMeans} & \cellcolor{yellow}\textbf{Affinity} & \textbf{Agglomerative} & \cellcolor{red}\textbf{MeanShift} & \cellcolor{orange}\textbf{GaussianMixture} & \textbf{HDBSCAN} & \textbf{Mean}    & \textbf{Max}     \\ \hline
\textbf{sec2vecText}                & 43.0   & 62.1     & 41.3          & 7.1       & 43.5            & 60.5    & 42.9     & 62.1    \\ \hline
\textbf{secText\_tfidf}             & 53.2   & 79.6     & 57.8          & 7.1       & 56.4             & 27.3    & 46.9    & 79.6    \\ \hline
\textbf{sec2vecMath\_op}            & 24.6   & 58.2     & 24.3          & 97.3      & 22.5            & 7.1     & 39.0    & 97.3    \\ \hline
\textbf{secMath\_op\_tfidf}         & 19.6   & 94.0     & 21.3          & 82.9      & 26.2            & 27.3    & 45.2    & 94.0    \\ \hline
\textbf{sec2vecMath\_id}            & 27.7   & 42.3     & 28.7          & 53.6      & 81.7            & 7.1     & 40.1    & 81.7    \\ \hline
\textbf{secMath\_id\_tfidf}         & 25.7   & 69.2     & 24.1          & 7.1       & 20.3            & 34.8    & 30.2     & 69.2    \\ \hline
\textbf{sec2vecMath\_opid}          & 33.2   & 41.9     & 32.5          & 53.6      & 57.5            & 7.1     & 37.6    & 57.5    \\ \hline
\textbf{secMath\_opid\_tfidf}       & 18.6   & 54.3     & 17.4          & 7.1       & 26.8            & 34.7    & 26.5     & 54.3    \\ \hline
\cellcolor{orange}\textbf{sec2vecMath\_surroundings}     & 51.7   & 61.6     & 30.6          & 98.2      & 61.2            & 7.1     & \cellcolor{yellow}51.7    & \cellcolor{red}98.2    \\ \hline
\textbf{secMath\_surroundings\_tfidf}  & 46.5   & 21.8     & 46.2          & 7.1       & 44.2             & 36.8    & 33.8    & 46.5    \\ \hline
\textbf{sec2vecTextMath\_opid}      & 43.4   & 68.7     & 41.8          & 7.1       & 41.4            & 40.9    & 40.6     & 68.7    \\ \hline
\textbf{sec2vecTextMath\_surroundings} & 47.0   & 62.8     & 40.3          & 86.7      & 65.0            & 7.1     & 51.5    & 86.7    \\ \hline
\textbf{Mean}                       & 36.2   & \cellcolor{yellow}59.7     & 33.9          & 42.9      & \cellcolor{yellow}45.6            & 24.8    & 40.5    & 59.7    \\ \hline
\textbf{Max}                        & 53.2   & 94.0     & 57.8          & \cellcolor{red}98.2      & \cellcolor{red}81.7            & 60.5    & 74.2    & 98.2    \\ \hline
\textbf{Runtime {[}\%{]}}          & 8.2    & \cellcolor{green}3.0      & 6.2           & 100.0     & 6.8             & 32.5    & 26.1   & 100.0   \\ \hline
\end{tabular}}
\\[30pt]
\resizebox{\textwidth}{!}{
\begin{tabular}{|l|l|l|l|l|l|l|l|l|}
\hline
\textbf{Encoding /Clusterer} & \textbf{KMeans} & \cellcolor{yellow}\textbf{Affinity} & \textbf{Agglomerative} & \cellcolor{red}\textbf{MeanShift} & \cellcolor{orange}\textbf{GaussianMixture} & \textbf{HDBSCAN} & \textbf{Mean}  & \textbf{Max}   \\ \hline
\textbf{abs2vecText}          & 37.5   & 75.0     & 31.8          & 98.3      & 40.6             & 7.1     & 48.4  & 98.3  \\ \hline
\textbf{absText\_tfidf}       & 32.1   & 58.9     & 53.6          & 7.1       & 25,1             & 70.2    & 41,2  & 70.2  \\ \hline
\cellcolor{orange}\textbf{abs2vecMath\_opid}    & 35.8   & 90.3     & 23.0          & 98.9      & 63.8            & 35.4    & \cellcolor{yellow}57.9  & \cellcolor{red}98.9  \\ \hline
\textbf{absMath\_opid\_tfidf} & 35.8   & 81.5     & 35.7          & 90.3      & 42.2            & 34.9    & 53.4  & 90.3  \\ \hline
\textbf{Mean}                 & 35.3   & \cellcolor{yellow}76.4     & 36.0          & 73.7      & \cellcolor{yellow}42,9            & 36.9    & 50,2  & 76.4  \\ \hline
\textbf{Max}                  & 37.5   & 90.3     & 53.6          & \cellcolor{red}98.9      & \cellcolor{red}63.8            & 70.2    & 69.1  & 98.9  \\ \hline
\textbf{Runtime {[}\%{]}}    & 3.0    & 2.0      & 1.1           & 100.0     & 3.6             & \cellcolor{green}0.4     & 18.3  & 100.0 \\ \hline
\end{tabular}}
\\[20pt]
\label{tab:Clust}
\end{table*}

Table \ref{tab:Clust} shows the cluster purities of the individual clustering algorithms using text or math encodings of the documents, sections, and abstracts.
The best encodings are doc/sec2vecMath\_surroundings and abs2vecMath\_opid. The most accurate algorithms are 1) GaussianMixture (highest mean and maximum), MeanShift (highest maximum), Affinity (highest max and mean). GaussianMixture is the best algorithm with a fixed cluster number, while Affinity and MeanShift are the best algorithms without a fixed cluster number. The fastest algorithm is HDBSCAN. Only for the mean of docs, text yields the highest value.
For the other mean and maximum values, math are better than text encodings.

A comparison of x2vec and tf-idf shows that the former outperforms the latter with mean(doc2vecX, sec2vecX, abs2vecX) = (56.3, 43.4, 53.1) $>$ mean(doc\_tfidf, sec\_tfidf, abs\_tfidf) = (51.5, 36.5, 47.3), and mean(2vec) = 50.9 $>$ mean(tfidf) = 45.0 summarized.
For abstracts, the math encodings yield better results than the text encodings with mean(abs\_text,abs\_math) = (44.8, 55.6).
However, given that mean(text, math, textmath) = (51.2, 48.2, 52.0), all in all, also for the clustering, the text encodings yield better results than the math encodings, but the combination of text and math slightly outperforms the other.

We tested some other algorithms that are not listed due to poor performance or exceedingly large runtimes (e.g. Spectral Clustering, DBSCAN).

\subsection{Human Expert vs. Computer Classification}

We carried out a human expert\footnote{We chose one physicist, able to separate all subject classes.} classification of 10 examples from each of the 14 subject classes for comparison to our algorithmic results. From 140 in total, 85 - i.e. 60.7\% were correctly classified.

\begin{figure*}[h]
\includegraphics[width=0.49\textwidth]{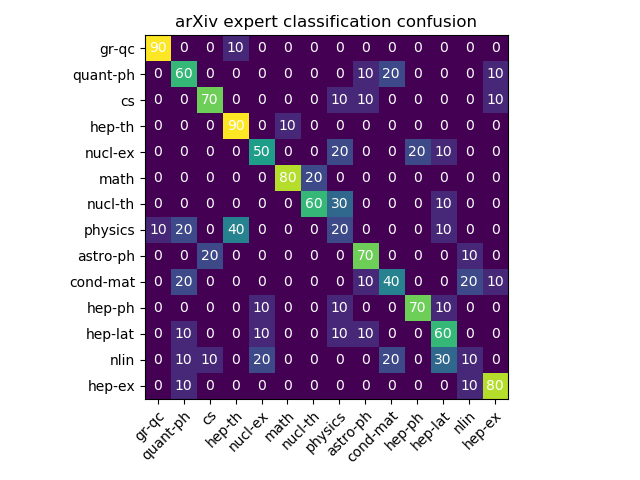}
\includegraphics[width=0.49\textwidth]{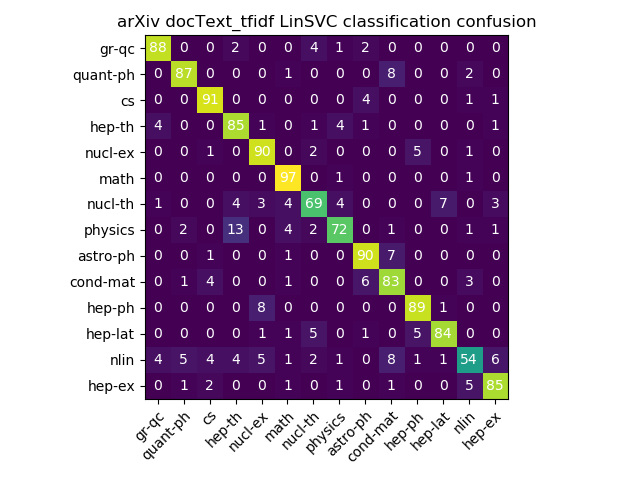}
\caption{Confusion matrix with percentages comparing the classification of a human expert (left) to the best performing combination of a LinSVC classifier on docText\_tfidf encodings (right).}
\label{fig:confusion}
\end{figure*}

Figure \ref{fig:confusion} shows a comparison of the classification confusion matrix. The human classifier (left) is outperformed by the computer classifier (right) with lower diagonal and higher off-diagonal values.

Both the computer and human classification confusion show that some categories like 'physics' should be disposed and distributed to the respective specializations ('cond-mat', 'hep-ex', 'nucl-ex', etc.).

\subsection{Multi-label Classification}

For the NTCIR and SigMathLing arxiv datasets, we could not find a baseline for our experiments. However, we were able to reproduce the results using the script of Schöneberg et al.~\cite{DBLP:conf/mkm/SchonebergS14} on 942337 ($\approx$ 1M) tf-idf encoded abstracts from the zbMath database~\cite{zbMATH}. Employing a multi-label Logistic Regression (LogReg) classifier on the 63 top level Mathematical Subject Classes (MSC), we could outperform the 53\% classification accuracy of the baseline ~\cite{DBLP:conf/mkm/SchonebergS14} by 27\% (to 80\%) for the prediction of the first two top level labels.

Furthermore, we tested the effect of the number of predicted labels on the classification accuracy to find a strong decrease predicting more labels.

The results show that predicting more than three labels per abstract does not yield reasonable results. This is why we rather concentrated on single-label classification.

\section{Conclusion and Outlook} \label{sec:ConclOutl}

\subsection{Summary}

In this paper, we discussed how methods of Natural Language Processing (NLP) and Mathematical Language Processing (MLP) can be combined to enable the use of Machine Learning (ML) in Information Retrieval (IR) applications on documents with mathematical content. We first provided a short review of MathIR, NLP and MLP and the current state of research in text and math classification and clustering. Subsequently, we introduced the employed datasets of mathematical documents and described encodings for their text and math content. We investigated the correlation between text and math similarity. Finally, we presented and discussed the results of a classification and clustering of 4900 documents, 3500 sections, and 1400 abstracts from the arXiv preprint server (\url{arxiv.org}).

The correlations between text and math (cosine) similarity (Table \ref{tab:Corr}) were relatively low (mean = 0.17, max = 0.27), motivating us to treat text and math encodings as separate features of a document.
While for the classification, the Text\_tfidf encoding was outperforming the others, for the clustering doc/sec2vecMath\_surroundings and abs2vecMath\_opid encodings are the best.
For both classification and clustering, the x2vec encodings yielded better results than the tf-idf encodings and text outperformed math encodings.
However, for the clustering, the combination of text and math slightly outperforms the separate encodings.
\\[0.25cm]
All in all, our \textbf{research questions} were answered as:

1) \textit{Combining text and math encodings does not improve the classification accuracy, but partly the cluster purity of selected ML algorithms working on documents, sections, and abstracts.}

2) \textit{On the whole, the doc2vec encoding outperforms tf-idf encoding. The most accurate classification algorithm is a Multilayer Perceptron (MLP), while for the clustering, the highest maximum and mean values of the purities are divided among GaussianMixture, MeanShift, and Affinity Propagation. The fastest algorithms are k-Nearest Neighbors, and Random Forest classifiers, and HDBSCAN clustering.}

\subsection{Discussion}

Why did the use of mathematical encodings not significantly improve classification accuracy? We suspect a low inter-class variance of the math encodings due to a large overlap of the formula identifier namespaces. For example, the identifier $x$ occurs very often in many subject classes, but with different meanings. Documents from different subject classes often have similar sets of identifier symbols. Therefore, we expect that disambiguation of the identifier semantics by annotation would increase the vector distance between subject classes and possibly increase classification accuracy. There are two ways to tackle the identifier disambiguation. It can be done supervised with or unsupervised without the quality control of a human. In the following, we present our results of unsupervised semantification using three different sources. Furthermore, we shortly discuss our ongoing endeavors to additionally perform supervised annotation in the future work section.

\subsection{Unsupervised Formula Semantification}

In an attempt to increase the classification accuracy compared to the previously presented math-encodings, we tested a conversion from math (identifier) symbols to text (semantics). We semantically enriched the text of the $14 \times 350 = 4900$ documents from the \textit{SigMathLing arXMLiv-08-2018} dataset by identifier name candidates provided from three different lists. These were previously extracted from the following sources:

\textbf{1) arXiv:} Identifier candidate names for all lower- and upper-case Latin and Greek letter identifier symbols appearing in the NTCIR arXiv corpus\footnote{http://ntcir-math.nii.ac.jp/data/} that was created as part of the NTCIR MathIR Task~\cite{DBLP:conf/ntcir/AizawaKOS14}. The candidates were extracted from the surrounding text of 60 M formulae and ranked by the frequency of their occurrence;

\textbf{2) Wikipedia:} Identifier candidate names extracted from definitions in mathematical English articles, as provided by \textit{Physikerwelt}\footnote{\url{https://en.wikipedia.org/wiki/User:Physikerwelt}};

\textbf{3) Wikidata:} Identifier candidate names retrieved via a SPARQL query\footnote{\url{https://query.wikidata.org}} for items with defining formula containing the respective identifier symbol.

For each source, the candidates were extracted, ranked by the occurrence frequency of the respective identifier symbol/name mapping, and dumped to static lists.
The encodings with semantic enrichment by the top 3 ranked identifier name candidates outperform all other mathematical encodings listed in Table \ref{tab:Class}. We will discuss an extension of the experiment to supervised semantification in our future work.

\subsection{Formula Encoding Challenge}

In this paper, we presented classification and clustering baselines on the \textit{SigMathLing arXMLiv-08-2018} and \textit{NTCIR-11/12 MathIR arXiv} datasets. Since, so far we were not able to significantly outperform the text encodings by math encodings, we call out for a \textit{Formula Encoding Challenge}. The aim is to find a suitable math encoding that outperforms or enhances the text classification.

\subsection{Future Work}

We now outline some future directions and experiments.

\paragraph{Deep Contextualized Encodings}

In the near future, we aim to test other recently developed encodings like Deep Bidirectional Transformers (BERT)~\cite{DBLP:journals/corr/abs-1810-04805} and Deep Contextualized Word Representations (ELMo)~\cite{DBLP:conf/naacl/PetersNIGCLZ18}, which are computationally more expensive and memory consuming. Our most extensive selection of text from 4900 documents (docText) taken from the SigMathLing arXMLiv-08-2018 dataset contains 1 million sentences, 75 million words, and 1.65 billion tokens and is thus larger than other NLP benchmark datasets the encodings are usually tested on. As an example for ELMo, the Stanford Natural Language Inference (SNLI) Corpus~\cite{DBLP:conf/emnlp/BowmanAPM15} comprises 570 thousand sentences, and the CoNLL-2003 Shared NER Task~\cite{DBLP:conf/conll/SangM03} unannotated data consists of 17 million tokens.

\paragraph{Supervised Formula Semantification}

To improve the classification accuracy on unsupervised semantification encodings, we plan to employ supervised semantification by human labeling. However, the semantic enrichment by "manual" annotation will take a significant amout of time. To facilitate and speed up the process, we are currently working on a formula and identifier name annotation recommender system~\cite{DBLP:conf/recsys/ScharpfMSBBG19}. We aim to integrate the tool into the editing views of both Wikipedia (Wikitext documents) and overleaf (LaTeX documents) to integrate the mathematical research community in the semantification process.

\paragraph{Formula Clustering}

We propose another potential use case of formula clustering, namely a \textit{formula occurrence retrieval}. Given a large dataset of mathematical documents (e.g., from the arXiv), the task will be to retrieve a ranking of formulae that occur most often. One could hypothesize that due to their popularity in research, the highest-scored formulae are most relevant candidates to have their underlying mathematical concepts seeded into encyclopedias and dictionaries such as Wikipedia, the semantic knowledge-base Wikidata~\cite{DBLP:journals/cacm/VrandecicK14} or the NIST Digital Library of Mathematical Functions~\cite{DBLP:journals/amai/Lozier03}. Since formulae often appear in a variety of different formulations or equivalent representations and it is a priori unknown how many different formula concepts~\cite{DBLP:conf/sigir/ScharpfSG18,DBLP:conf/sigir/ScharpfSCG19} will be discovered (the cluster parameter $k$), this is a very challenging problem that the authors currently are working on.

\section*{Acknowledgment}

This work was supported by the German Research Foundation (DFG grant GI-1259-1).
The authors would like to thank Christian Borgelt for his support.

\printbibliography[keyword=primary]

\end{document}